\documentclass[preprintnumbers, prd, floatfix, letterpaper, superscriptaddress,nofootinbib]{revtex4-1}
\usepackage{amsmath}
\usepackage{amsfonts}
\usepackage{amssymb}
\usepackage{mathtools}
\usepackage{bm}
\usepackage{xcolor}
\usepackage{multirow} 
\usepackage{xcolor}
\usepackage{graphicx}
\usepackage{caption}
\usepackage{comment}
\usepackage[T1]{fontenc}
\usepackage[pdftex, pdftitle={Article}, pdfauthor={Author}]{hyperref} 
\usepackage{scalerel}
\usepackage{tikz}
\usetikzlibrary{svg.path}

\definecolor{orcidlogocol}{HTML}{A6CE39}
\tikzset{
  orcidlogo/.pic={
    \fill[orcidlogocol] svg{M256,128c0,70.7-57.3,128-128,128C57.3,256,0,198.7,0,128C0,57.3,57.3,0,128,0C198.7,0,256,57.3,256,128z};
    \fill[white] svg{M86.3,186.2H70.9V79.1h15.4v48.4V186.2z}
                 svg{M108.9,79.1h41.6c39.6,0,57,28.3,57,53.6c0,27.5-21.5,53.6-56.8,53.6h-41.8V79.1z M124.3,172.4h24.5c34.9,0,42.9-26.5,42.9-39.7c0-21.5-13.7-39.7-43.7-39.7h-23.7V172.4z}
                 svg{M88.7,56.8c0,5.5-4.5,10.1-10.1,10.1c-5.6,0-10.1-4.6-10.1-10.1c0-5.6,4.5-10.1,10.1-10.1C84.2,46.7,88.7,51.3,88.7,56.8z};
  }
}

\newcommand\orcidicon[1]{\href{https://orcid.org/#1}{\mbox{\scalerel*{
\begin{tikzpicture}[yscale=-1,transform shape]
\pic{orcidlogo};
\end{tikzpicture}
}{|}}}}

\begin{document}
\title{First-order phase transitions and cosmic evolution: thermodynamic approach to generalized holographic dark energy}
\author{Miguel Cruz\orcidicon{0000-0003-3826-1321}}
\email{miguelcruz02@uv.mx}
\affiliation{Facultad de F\'{\i}sica, Universidad Veracruzana 91097, Xalapa, Veracruz, M\'exico,}

\author{Joaquín Housset\orcidicon{0009-0002-4225-2102}}
\email{joaquin.housset@pucv.cl}

\author{Samuel Lepe\orcidicon{0000-0002-3464-8337}}
\email{samuel.lepe@pucv.cl}
\author{Joel Saavedra\orcidicon{0000-0002-1430-3008}}
\email{joel.saavedra@pucv.cl}
\affiliation{Instituto de Física, Pontificia Universidad Católica de Valparaíso, Casilla 4950, Valparaíso, Chile,}

\author{Francisco Tello-Ortiz\orcidicon{0000-0002-7104-5746}}
\email{francisco.tello@ufrontera.cl}
\affiliation{Departamento de Ciencias Físicas, Universidad de La Frontera, Casilla 54-D, 4811186 Temuco, Chile.}

\date{\today}

\begin{abstract}
Focusing on the description of cosmic evolution at late times, this study examines a generalized holographic dark energy (HDE) framework constructed via a polynomial expansion in the Hubble parameter, which includes contributions proportional to $H^{2}$, $H^{4}$, and $H^{6}$, introduced through a variable parameter within the standard holographic formula. The analysis is carried out in the context of a spatially flat Friedmann–Lemaître–Robertson–Walker (FLRW) Universe, consisting of non-interacting matter together with the HDE fluid. We obtain the full set of Friedmann equations to investigate cosmic evolution and then analyze the system to determine whether thermodynamic $P - v$ type phase transitions can occur.
\end{abstract}

\maketitle

\section{Introduction}
Within the thermodynamic description of the Universe under Einstein gravity, the possibility of Van der Waals-type phase transitions is governed by the geometry of the apparent horizon. This geometric constraint, in turn, requires a variable dark energy sector. This intrinsic link, as established in \cite{cruzlepe}, imposes a crucial requirement: some cosmological quantities must be expressed in terms of the apparent horizon radius, denoted as $R_A$ in this work, and given by \cite{helo0, faraoni}\footnote{It is well-known that if we define the following metric, $h_{ab} := \mbox{diag}[-1,a^{2}(t)/(1-kr^{2})]$, with $a,b=t,r$, the usual FLRW spacetime metric $ds^{2} = -dt^{2}+a^{2}(t)\left[\frac{dr^{2}}{1-kr^{2}}+r^{2}(d\theta^{2}+\sin^{2}\theta d\varphi^{2}) \right]$ with spatial curvature $k = 0, \pm 1$, takes the form, $ds^{2} = h_{ab}dx^{a}dx^{b} + R^{2}(t)(d\theta^{2}+\sin^{2}\theta d\varphi^{2})$, where $R(t,r)$ is the physical radius of the FLRW Universe, $R(t,r) \equiv a(t)r$, with $r$ being the comoving coordinate and $a(t)$ the scale factor. Then by solving the condition $ h^{ab}\partial_{a}R \partial_{b}R =0$, we arrive to an expression for $R_{A}$ given above.} $R_{A} = 1/\sqrt{H^{2}+\frac{k}{a^{2}}}$.
In the Einstein gravity scenario, a single-fluid description of the Universe is insufficient to produce critical phenomena. It is instead necessary to decompose the total energy content into its two primary contributions: dark energy and dark matter; and relate them through the cosmological coincidence parameter. HDE naturally satisfies the requirement of linking the geometry of the apparent horizon to the thermodynamical description of the Universe, positioning it as a strong candidate for such analyzes. The use of this reformulation in standard cosmology is crucial to admit critical phenomena in the equation of state of the Universe. The compelling feature of HDE is its foundation in the holographic principle, which emerges from black hole thermodynamics. This principle constrains the total energy of a system by its surface area (a UV bound) and, when applied to cosmology, necessitates the existence of an infrared (IR) cutoff that defines the maximal observable scale. Consequently, the HDE model is endowed with a strong physical justification rooted in quantum gravity considerations, distinguishing it as a well-motivated theoretical construction rather than an ad hoc phenomenological fit to observational data. 

Mathematically, the holographic dark energy model is specified by the inequality $\rho_{\mathrm{de}}L^{4}\leq S$ where $L$ is the length of the horizon and $S$ its associated entropy where the usual assumption is considered for the entropy, $S \propto A \propto L^{2}$ with $A$ being the area \cite{bekenstein, haw}, therefore we have for the dark energy density \cite{cohen, li, Shapiro, Wang, Malekjani}
\begin{equation}
    \rho_{\mathrm{de}}\,=\,3\,c^{2}\,M^{2}_{\mathrm{pl}}\,L^{-2},
\end{equation}
where the factor $3$ was introduced for convenience and $c^{2}$ is a dimensionless parameter that is usually assumed constant, $L$ corresponds to the characteristic length or cutoff. This equation is consistent with the holographic principle, establishing that the vacuum energy is bounded by the characteristic length $L$. Consequently, for a length scale of cosmological size, the resulting dark energy density is naturally small, a value that corresponds remarkably well to current observations. This is not merely a coincidence; it strongly suggests that the observed smallness of the cosmological constant may be a direct consequence of a fundamental physical principle linking the vacuum energy to the maximum length scale of the Universe. Following a dimensional analysis and invoking the holographic principle, dark energy admits the expansion given below, which depends only on two quantities: the reduced Planck mass, $M_{\mathrm{pl}}$, and the characteristic length $L$ \cite{Shapiro, Wang, Nojiri1, Nojiri2}
\begin{equation}
    \rho_{\mathrm{de}}= C_{1}M^{4}_{\mathrm{pl}}+C_{2}M^{2}_{\mathrm{pl}}L^{-2}+C_{3}L^{-4}+\mathcal{O}(L^{-6})+...
    \label{eq:expansion}
\end{equation}
where the parameters $C_{1}, C_{2}, C_{3}$ are constant. The constant term, $C_1$, is excluded from the expansion, as its inclusion would reintroduce the problem of cosmological constant value. Consequently, the leading order contribution to the energy density is the $L^{-2}$ term, while the $L^{-4}$ and $L^{-6}$ terms are interpreted as higher order corrections arising from high energy physics. Within this framework, the appropriate infrared cutoff $L$ is identified with the Hubble scale, $L\,=\,L\,(H)$\footnote{A generalized version of HDE was proposed in Ref. \cite{hdeodint}. In this approach, the infrared cutoff can include derivatives of both the particle and future horizons and may also depend on the scale factor.}, due to its direct relationship to the apparent horizon, as can be seen in the definition given above, notice that for the flat case ($k=0$) we have $R_{A}=H^{-1}$. This choice is also physically motivated by the fact that the Hubble horizon acts as a causal boundary, which can be endowed with well-defined thermodynamic properties, such as temperature and entropy \cite{cai, quevedo}. Using the Hubble scale as the characteristic length, $L\,=\,1\,/\,H$, allow us to write for the energy density  
\begin{equation}
    \rho_{\text{de}}\,=\,3\,c^{2}\,M^{2}_{\mathrm{pl}}\,H^{2}. \label{eq:li}
\end{equation}
The resulting proportionality between the energy density and the square of the Hubble parameter provides a natural explanation for the observed magnitude of dark energy \cite{li}. By linking its present value directly with the current expansion rate, $H_0$, the model accounts for this quantity without fine-tuning. Furthermore, this framework inherently predicts a dynamical dark energy sector, whose density evolves throughout the history of the Universe\footnote{An intriguing feature of the holographic framework is that it can account for both the early-time and late-time accelerated expansion phases of our universe within a single, unified description, as shown in Ref. \cite{hdeinfla}; see also \cite{hdeinfla2}.}. Despite this interesting property of the HDE model, Hsu pointed out in \cite{Hsu:2004ri} that its associated equation of state over cosmological distances is not consistent and, consequently, is strongly disfavored by observational data. To avoid this issue, a simple generalization for the energy density (\ref{eq:li}) was proposed and consisted of considering a variable parameter $c^{2}$ written as a function of time instead of a constant value. However, for consistency $c^{2}(t)$ must vary slowly, this means that the quotient $\dot{c}^{2}(t)/c^{2}(t)$ is upper bounded by the expansion rate \cite{pavon1, pavon2, slow, Zimdahl, Radicella}, i.e.,
\begin{equation}
    \frac{\dot{c}^{2}(t)}{c^{2}(t)} \leq H(t). \label{eq:slow}
\end{equation}
This latter condition must be fulfilled in order to describe in a consistent way the thermal evolution of the Universe, allowing a transition from decelerated to accelerated cosmic expansion. The present study is dedicated to the cosmological and thermodynamic analysis of a generalized HDE model. The framework is based on the identification of the Hubble scale as the characteristic length and extends the standard $H^{2}$ paradigm to include high energy physics corrections ($H^{4}$ and $H^{6}$). These corrections are introduced into the HDE density via a controlling variable parameter, and the resultant implications for cosmic evolution and thermodynamics are thoroughly investigated. In cosmological models, higher order curvature corrections\footnote{In the high energy regime, a primary motivation for these higher curvature terms is based on their improvement results of the renormalization properties of general relativity \cite{stelle}.} to general relativity naturally give rise to corrective terms in the cosmic acceleration which involve powers of the Hubble parameter. Notably, these corrections can be formulated such that the resulting field equations remain second order; see, for instance, \cite{cubic, sixth, oliva}. Our study is performed at background level in a flat FLRW Universe described by the line element 
\begin{equation}
ds^2 = -dt^2 + a^2(t)\left[dr^2 + r^2 d\Omega^2\right],
\end{equation}
where $d\Omega^{2}=d\theta^{2}+\sin^{2}\theta d\varphi^{2}$. The work is organized as follows: Section \ref{sec:2} formalizes the Generalized HDE model and derives the set of modified Friedmann equations governing the cosmic evolution. Section \ref{sec:3} is dedicated to exploring the thermodynamic properties of the system, specifically analyzing the conditions required for the existence of $P-v$ type phase transitions. In Section \ref{sec:4}, we investigate the cosmological evolution of the model and relate it to the thermodynamic findings. Finally, Section \ref{sec:5} presents the general conclusions drawn from this investigation and discusses avenues for future research. From now on, we adopt natural units where $G=c=M_{\rm{pl}}=1$.

\section{Holographic energy density and dynamics}
\label{sec:2}

As commented previously, the allowed energy density in Quantum Field Theory (QFT) is naturally arranged into even powers of $H$ with the coefficients suppressed by UV scaling \cite{Shapiro}, the $H^{2}$ term is the late time leader, and high energy corrections come as $H^{4}$ and $H^{6}$ terms. We propose the HDE density as:
\begin{equation}\rho_{\mathrm{de}}=3c^{2}(t)H^{2}(t), \label{eq:hdevar}
\end{equation}
where we have defined
\begin{equation}
    c^{2}(t) = \beta_{1} + \beta_{2}H^{2}(t) + \beta_{3}H^{4}(t).\label{eq:coeff}
\end{equation}
So, the generalized HDE density is taken as the sum of the first three powers of $H$ with $\beta_{1}$, $\beta_{2}$ and $\beta_{3}$ as free parameters of the model. 
Therefore, Eq. (\ref{eq:hdevar}) must be understood as an effective holographic closure relation, rather than as a fundamental microscopic definition of the energy-momentum tensor. In our cosmological model, the horizon-based thermodynamic framework for the dark energy sector treats it as an emergent gravitational component whose energy density is determined by geometric quantities associated with the apparent horizon. In this way, the dependence of $\rho_{\rm de}$ on the Hubble rate does not imply a reversal of dynamical causality; instead it reflects the nonlocal nature of the gravitational degrees of freedom. The Friedmann equation is satisfied throughout and serves as a consistency condition that selects the physically admissible solutions. In particular, in the dark energy regime, the equation
\begin{equation}
3H^2 = \rho_{\rm de}(H),
\end{equation}
defines the allowed cosmological branches. We did not assume that the right-hand side represents a higher-order perturbative expansion around $3H^2$. 
Accordingly, the model should be interpreted as a nonlinear effective description, analogous to other geometric or horizon-based approaches to dark energy, rather than as a low-energy expansion with controlled truncation. Furthermore, as shown in Ref. \cite{cruzlepe}, a dark energy model described by powers of $H$ remains classically stable.

In the context of a flat FLRW Universe, the Friedmann equations take the form
\begin{equation}
    3H^{2} = 8\pi (\rho_{\mathrm{m}}+\rho_{\mathrm{de}})=8\pi \rho_{\text{de}} (1+r),\label{eq:fried1}
\end{equation}
where the components of the Universe are characterized by a perfect fluid with energy density $\rho$ and pressure, $p$. In our notation $\rho_{\mathrm{m}}$ characterizes the energy density of dark matter and $\rho_{\mathrm{de}}$ is associated with the dark energy sector. We have defined the cosmological coincidence parameter $r$ as the quotient $r:=\rho_{\mathrm{m}}/\rho_{\mathrm{de}}$ to relate the noninteracting fluids. As examined in \cite{cruzlepe}, within the framework of standard cosmology, it is essential to consider two dominant cosmic fluids to reveal the critical phenomena at the thermodynamic level. Rather than resorting to modified gravity theories, this method relies solely on a dynamical dark energy component, which simultaneously determines the structure of the coincidence parameter and generalizes the effective pressure expression, as we also show below. The acceleration equation is written in the usual form
\begin{equation}
    \dot{H} = -4\pi (\rho_{\mathrm{m}}+\rho_{\mathrm{de}}+p_{\mathrm{m}}+p_{\mathrm{de}}). \label{eq:fried2}
\end{equation}

\subsection{Apparent horizon thermodynamics and equation of state}

In a spatially flat FLRW Universe, the apparent horizon plays the role of a causal boundary with radius
\begin{equation}
R_{A}=\frac{1}{H}.
\end{equation}
Since the apparent horizon can be considered a thermodynamic system, we can compute the work density of the matter fields as
\begin{equation}
    W=-\frac{1}{2}T = \frac{1}{2}(\rho-p),\label{eq:work}
\end{equation}
being $T$ the two-dimensional normal trace of the energy-momentum tensor, $T:=h^{ab}T_{ab}$, with $T_{ab} = (\rho+p)u_{a}u_{b}+pg_{ab}$. Due to the interpretation of thermodynamic system for the apparent horizon, we can associate surface gravity with it; in this case, we have
\begin{equation}
    \kappa = \frac{1}{2\sqrt{-h}}\partial_{a}(\sqrt{-h}h^{ab}\partial_{b}R),
\end{equation}
yielding
\begin{equation}
    \kappa = -\frac{R}{2}\left[\dot{H}+2H^{2}+\frac{k}{a^{2}} \right].
\end{equation}
Then, if we evaluate at $R=R_{A}$, the surface gravity takes the form
\begin{equation}
    \kappa = -\frac{1}{R_{A}}\left(1-\frac{\dot{R}_{A}}{2HR_{A}} \right),
\end{equation}
which, in terms of the cosmographic parameter $q:=-1-\dot{H}/H^{2}$, can be expressed as \cite{cruzlepe}
\begin{equation}
    \kappa = -\frac{1}{2R_{A}}(1-q).
\end{equation}
This latter expression indicates that, for an expanding Universe, we have $\kappa < 0$ since $q<0$. Formally, the sign of the temperature associated with the apparent horizon depends on the nature of the horizon. As discussed in Refs. \cite{helou,posit}, the expanding cosmology corresponds to a past-inner trapping horizon. In this case, we must have: $T_{A} \propto -\kappa$ and $\kappa < 0$, as commented above, for the surface gravity. These conditions lead to a positive physical temperature given as
\begin{equation}
    T_{A} = -\frac{\kappa}{2\pi} = \frac{1}{2\pi R_{A}}\left(1-\frac{\dot{R}_{A}}{2HR_{A}}\right) = \frac{1}{4\pi R_{A}}(1-q). \label{eq:temp}
\end{equation}
Notice that the horizon temperature given by $T_{A}=H/2\pi$ corresponds to the case $q=-1$. The amount of energy inside the horizon is simply
\begin{equation}
    E=\rho V_{A} = \frac{4\pi}{3}\rho R^{3}_{A},
\end{equation}
with $V_{A}$ being the volume enclosed by the apparent horizon. Then, the differential for the energy can be computed straightforwardly
\begin{equation}
    dE = \frac{4\pi}{3}R^{3}_{A}d\rho + 4\pi \rho R^{2}_{A}dR_{A}.
\end{equation}
This leads to the following expression after using the previously mentioned results for the apparent horizon:
\begin{equation}
  dE = \frac{4\pi}{3}R^{3}_{A}d\rho + 2\pi R^{2}_{A}(\rho + p)dR_{A}+WdV_{A}, \label{eq:de}
\end{equation}
which remains valid for any cosmological model \cite{law}. Notice that, according to Eq. (\ref{eq:de}), $W$ can be related to the thermodynamic pressure, $P$, as it is the conjugate variable of the thermodynamic volume. On the other hand, from the Friedmann constraint (\ref{eq:fried1}) and the energy density (\ref{eq:hdevar}), it is direct to obtain the coincidence parameter as a function of the radius of the apparent horizon through the coefficient $c^{2}$, resulting in the following expression
\begin{equation}
1+r(R_{A})=\frac{1}{8\pi\,c^{2}(R_A)},\label{eq:coinci}
\end{equation}
which in turn results in a relevant expression for our purposes as we shall see below.

\subsection{Thermodynamic pressure and equation of state}

Applying the unified first law at the apparent horizon we have $W\equiv P$, the effective thermodynamic pressure is obtained from (\ref{eq:work}) and the Friedmann equations (\ref{eq:fried1}) and (\ref{eq:fried2}), then we can write the equation of state as \cite{cruzlepe} 
\begin{equation}
P(R_{A},T_{A},r)=\frac{T_{A}}{2R_{A}}+\left(\frac{3}{1+r(R_{A})}-1\right)\frac{1}{16\pi R_{A}^{2}},
\end{equation}
since we are interested in the thermodynamic description of the cosmological model, it is convenient to introduce the specific volume defined as $v\equiv 2R_{A}$, then the pressure takes the following form
\begin{equation}
P(v,T_{A},c^{2})=\frac{T_{A}}{v}-\frac{1}{4\pi v^{2}}+\frac{6\,c^{2}(v)}{v^{2}},\label{eq:EoS_PvT}
\end{equation}
where we have considered Eq. (\ref{eq:coinci}). This latter expression provides a direct thermodynamic interpretation of the holographic dark sector, fully analogous to the $P-v$ equation of state of van der Waals fluids. Using the standard definition of classical thermodynamics, criticality occurs when \cite{callen}
\begin{equation}
\left(\frac{\partial P}{\partial v}\right)_{T_{A}}=0, \qquad 
\left(\frac{\partial^{2} P}{\partial v^{2}}\right)_{T_{A}}=0,
\end{equation}
leading to critical values $(P_{c},v_{c},T_{c})$ associated with first–order phase transitions of cosmological origin. As can be seen, the nontrivial dependence on $R_A$ of the coincidence parameter opens the door to critical phenomena within the Einstein framework. Consequently, as evident from Eq. (\ref{eq:EoS_PvT}), a significantly richer cosmological thermodynamics may arise from a minimal extension of the standard scenario, in which, for a single fluid description, the effective pressure constructed above is simply given by
\begin{equation}
    P(v,T_{A}) = \frac{T_{A}}{v}+\frac{1}{2\pi v^{2}},\label{eq:usual}
\end{equation}
showing no critical behavior, as is already known.

\section{Isotherms, critical points, and phase transition}
\label{sec:3}

Evaluating the derivatives of \eqref{eq:EoS_PvT} with the consideration of the explicit form given in Eq. (\ref{eq:coeff}) for the coefficient $c^{2}(v)$, allows us to write the following results
\begin{align}
& \frac{\partial P}{\partial v} = -\frac{T_{A}}{v^{2}}+\frac{1}{2\pi v^{3}}-\frac{12\beta_1}{v^{3}}-\frac{96\beta_2}{v^{5}}-\frac{576\beta_3}{v^{7}},\\
& \frac{\partial^{2}P}{\partial v^{2}} = \frac{2T_{A}}{v^{3}}-\frac{3}{2\pi v^{4}}
+\frac{36\beta_{1}}{v^{4}}+\frac{480\beta_{2}}{v^{6}}+\frac{4032\beta_3}{v^{8}}.
\end{align}
Eliminating $T_{A}$ from these conditions leads to a quadratic equation for $y=v_{c}^{2}$:
\begin{equation}
A\,y^{2}+144\,\beta_{2}\,y+1440\,\beta_{3}=0,
\label{eq:crit_poly}
\end{equation}
with $A:=1/(4\pi)-6\beta_{1}$, whose positive solution gives $v_{c}=\sqrt{y}$. The corresponding critical temperature and pressure follow as
\begin{align}
T_{c} &= -\left(\frac{2A}{v_{c}}+\frac{96\beta_{2}}{v_{c}^{3}}+\frac{576\beta_{3}}{v_{c}^{5}}\right), \\[2mm]
P_{c} &= \frac{T_{c}}{v_{c}}-\frac{1}{4\pi v_{c}^{2}}+\frac{6\beta_{1}}{v_{c}^{2}}
+\frac{24\beta_{2}}{v_{c}^{4}}+\frac{96\beta_{3}}{v_{c}^{6}}.
\end{align}
As will be explained in the next section, to guarantee physically viable solutions, the model parameters are strictly constrained, including a positive critical volume and a real, positive Hubble parameter. These constraints naturally result in a positive critical temperature ($T_{c}>0$). To maintain physical consistency, this same valid parameter set is used to generate the diagram $P\text{-}v$ and the Gibbs free energy plots presented below.

\subsection{Isotherms and phase transition}
\begin{figure}[h!]
\centering
\includegraphics[width=0.7\textwidth]{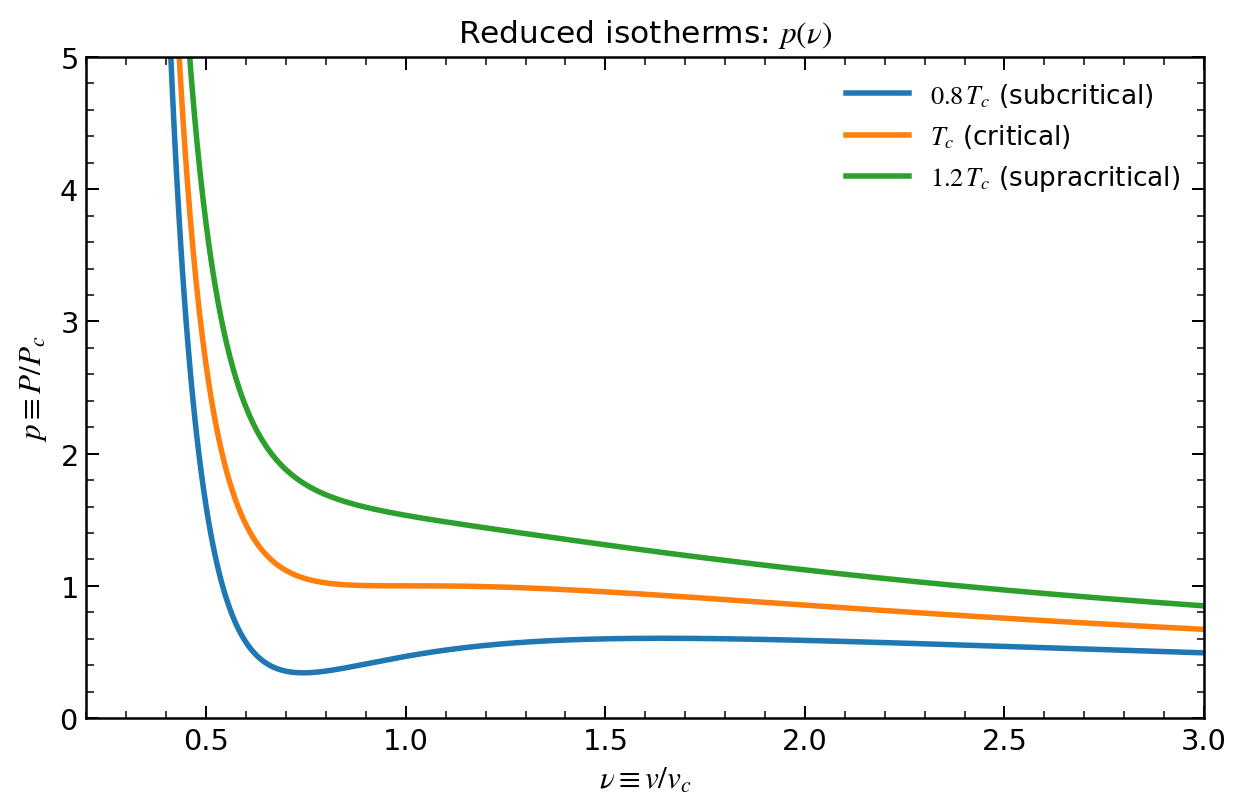}
\caption{Dimensionless $P - v$ isotherms for the generalized HDE model. In accordance with the qualitative behavior provided above, the orange curve represents the critical temperature $T_{c}$ isotherm with the critical point at the inflection point. The subcritical regime $T_{A}<T_{c}$ is represented by the blue curve and exhibits a coexistence of phases (two different values of volume for the same pressure). In the supercritical region ($T_{A}>T_{c}$, green curve), the isotherms are monotonic, indicating a single stable phase.}
\label{fig:isotherms}
\end{figure}

To have dimensionless quantities, we define the following reduced variables: $P/P_{c}$, $T/T_{c}$ and $v/v_{c}$. In Fig. (\ref{fig:isotherms}), we show the behavior of expression (\ref{eq:EoS_PvT}) in terms of the reduced variables. Qualitative behavior is as follows:
\begin{itemize}
\item For $T_{A}>T_{c}$ (supercritical region) the isotherms are monotonic and no phase transition occurs.
\item At $T_{A}=T_{c}$ the isotherm displays a point of inflection at $(v_{c},P_{c})$.
\item For $T_{A}<T_{c}$ (subcritical regime) the isotherms develop a non–monotonic loop, with an intermediate region of positive slope $\partial P/\partial v >0$ which signals thermodynamic instability. The physically realized transition is determined by the Maxwell equal–area construction, which replaces the oscillatory branch by a constant–pressure line connecting two coexistence volumes $v_{1}$ and $v_{2}$.
\end{itemize}

This analysis demonstrates that the generalized holographic model exhibits a thermodynamic phase structure closely analogous to van der Waals fluids, with a first–order phase transition below the critical temperature and a second–order critical point.

\subsection{Gibbs free energy and phase transition}

\begin{figure}[h!]
\centering
\includegraphics[width=0.7\textwidth]{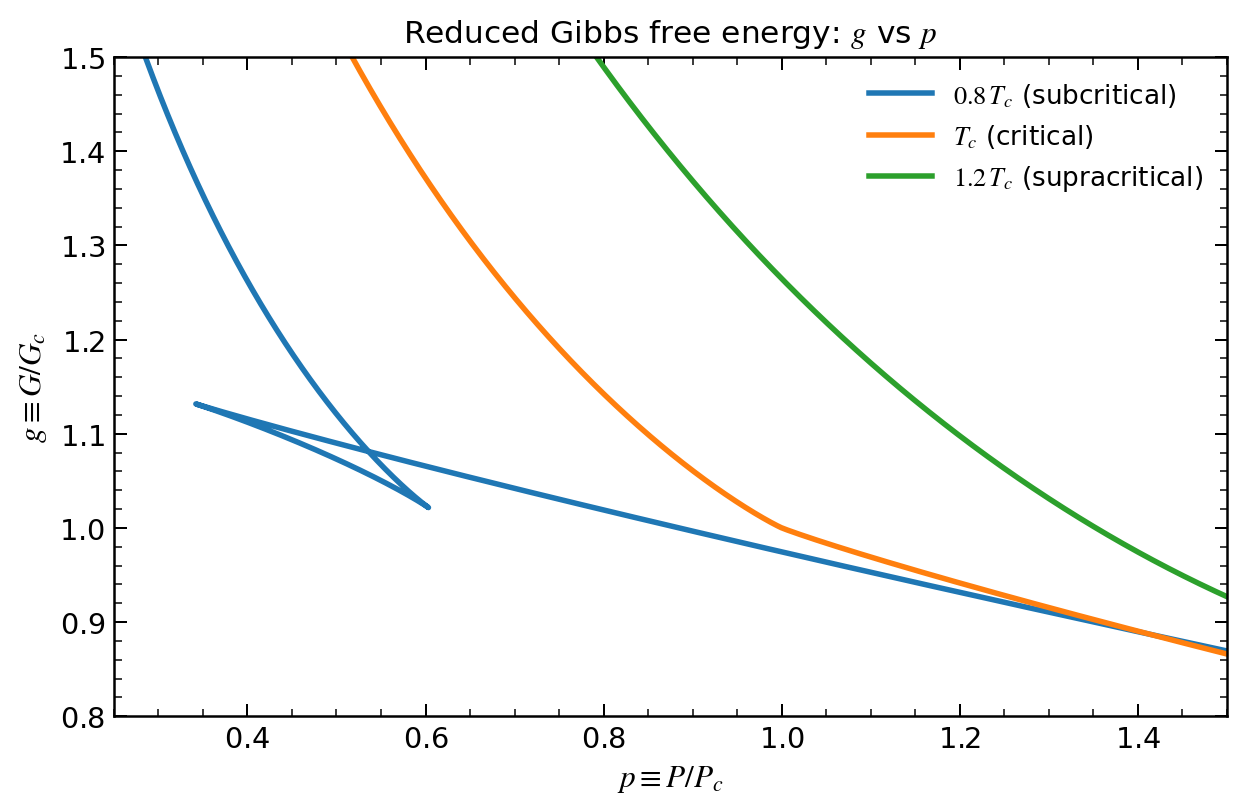}
\caption{Reduced Gibbs free energy as a function of pressure.  
Subcritical isotherm ($0.8T_c$) exhibit the characteristic swallowtail structure, indicating a first–order phase transition.    
For $T_A>T_c$ (shown here at $1.2T_c$) the swallowtail disappears and the Gibbs free energy becomes smooth and single-valued.}
\label{fig:gibbs}
\end{figure}

The thermodynamic behavior of the holographic model can be further clarified by considering the Gibbs free energy to corroborate the nature of the phase transition undergone by the system. From the equation of state \eqref{eq:EoS_PvT}, the reduced Gibbs free energy (up to an additive function of $T_A$ only) is obtained as
\begin{equation}
G(v,T_{A}) = T_{A}\!\left(1-\ln v\right) 
+ \frac{2A}{v} + \frac{32\beta_{2}}{v^{3}} 
+ \frac{576}{5}\frac{\beta_{3}}{v^{5}},
\label{eq:Gibbs}
\end{equation}
and its behavior is shown in Fig. (\ref{fig:gibbs}). At a given temperature $T_A$, the physical branch is obtained by minimizing $G$ at fixed pressure. For $T_A<T_c$ the Gibbs free energy displays the characteristic \emph{swallowtail} structure when plotted as a function of $P$, revealing the coexistence of two phases.  
The Maxwell construction ensures that the transition occurs at a constant pressure $P^\star$, where the Gibbs free energies of the two phases coincide.  
At $T_A=T_c$ the swallowtail shrinks to a cusp at $(P_c,G_c)$, and for $T_A>T_c$ the function $G(P)$ becomes single-valued and monotonic, signaling the disappearance of the first–order transition.

Therefore, this analysis confirms that the generalized holographic model supports a van der Waals type phase structure: a line of first–order transitions ending at a critical point, beyond which only a single homogeneous phase is present.

\section{Modified Friedmann dynamics and dark energy equation of state}
\label{sec:4}

\begin{figure}[h!]
\centering
\includegraphics[width=0.7\textwidth]{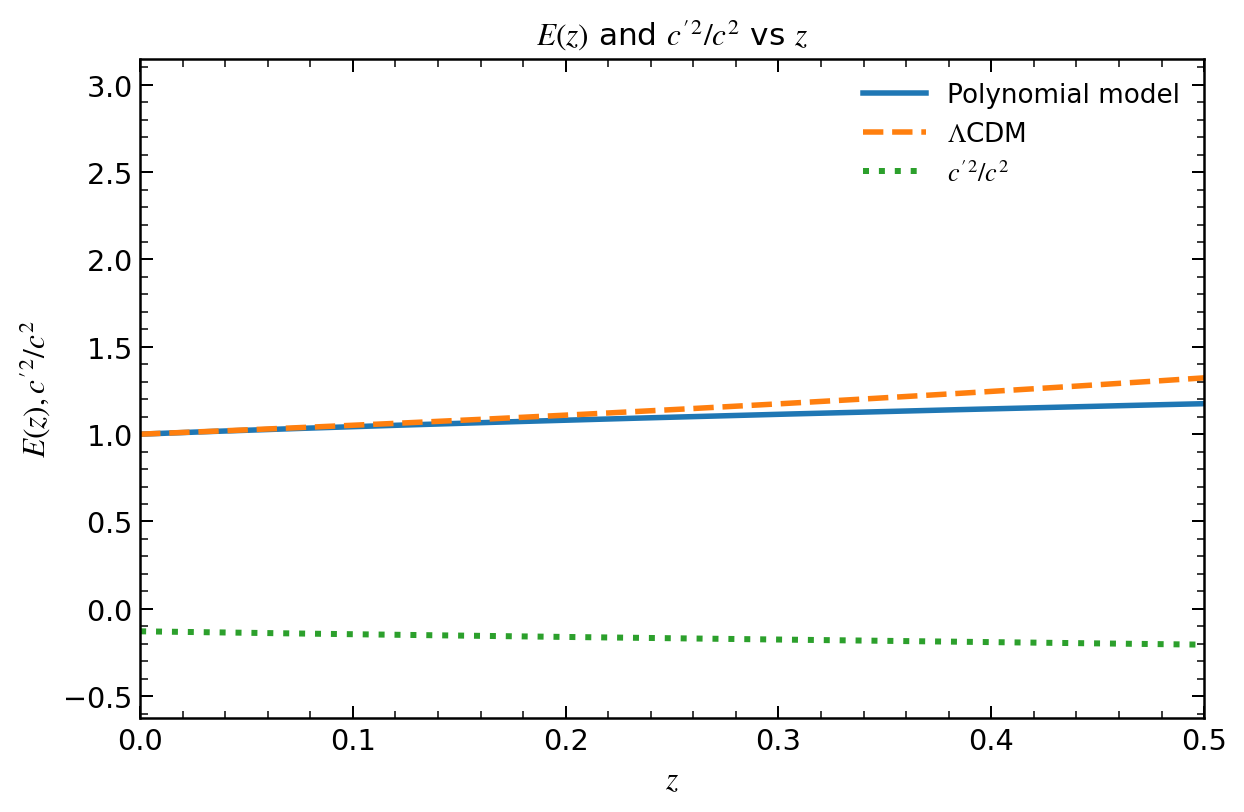}
\caption{$H(z)/H_{0}$ for the holographic model compared to flat $\Lambda$CDM with $\Omega_{m0}=0.315$. Also it's observed that the term $c'^{2}/c^{2}$ it remains bounded above by $H(z)$ over the entire range considered. In this case the prime denotes derivatives with respect to the cosmological redshift.}
\label{fig:Hzhde}
\end{figure}

To investigate the link between our model's dynamics and its thermodynamics, we analyze the evolution of the Hubble parameter, which incorporates the total contribution from the dark sector. We define the normalized Hubble parameter as $E(z) \equiv H(z)/H_0$, where $H_0 \equiv H(z=0)$ is the Hubble constant. The redshift $z$ is given by $1+z = 1/a(t)$, assuming a present-day scale factor normalized to unity ($a_0=1$). Subsequently, the zero subscript will consistently denote the values of the cosmological parameters at the present time ($z=0$). The normalized Hubble parameter induces the definition of the fractional energy density parameters, for the matter sector we have, $\Omega_{\mathrm{m,0}}=8\pi \rho_{\mathrm{m,0}}/(3H_{0}^{2})$ and considering the HDE (\ref{eq:hdevar}) together with (\ref{eq:coeff}) and the Friedmann constraint (\ref{eq:fried1}) one can define the following
\begin{equation}
\alpha_{1}=8\pi \beta_{1},\quad 
\alpha_{2}=8\pi \beta_{2}H_{0}^{2},\quad 
\alpha_{3}=8\pi \beta_{3}H_{0}^{4},
\end{equation}
therefore, Eq. (\ref{eq:fried1}) takes the form
\begin{equation}
\alpha_{3}E^{6}(z)+\alpha_{2}E^{4}(z)+(\alpha_{1}-1)E^{2}(z)+\Omega_{\mathrm{m,0}}(1+z)^{3}=0.
\end{equation}
If we consider the change of variable $X(z) \equiv E(z)^2$, the previous equation can be written as the standard cubic equation $ax^{3}+bx^{2}+cx+d=0$,
\begin{equation}
\alpha_3\,X^3 + \alpha_2\,X^2 + (\alpha_1-1)\,X + \Omega_{\mathrm{m,0}}(1+z)^3 = 0,
\label{eq:cubic}
\end{equation}
where we identify the coefficients $a=\alpha_3, b=\alpha_2, c=\alpha_1-1$ and $d=\Omega_{\mathrm{m,0}}(1+z)^3$. In this case, normalization at the present time $X(0)=1$ implies the following condition: $\alpha_1+\alpha_2+\alpha_3=1-\Omega_{\mathrm{m,0}}$. As discussed previously, to obtain a positive critical volume \(v_{c}\), the physically meaningful solutions of Eq. \eqref{eq:crit_poly}, which satisfy the condition \(v_{c}=\sqrt{y}\geq 0\), are permitted only when the coefficient \(A\) is non-negative. From its definition $A = 1/(4\pi)-6\beta_{1}\geq0$ it immediately follows that $\beta_1 \leq 1/(24\pi)$. This imposes a strict theoretical upper bound of $\beta_{1} \leq 1/(24\pi) \simeq 0.0132$. Therefore, we set the value $\beta_1 = 0.01$ in our analysis. By imposing the normalization condition together with the previously specified value of the parameter $\beta_{1}$, we determine the remaining parameters $\alpha_2$ and $\alpha_3$. We also adopt the values provided by the Planck collaboration \cite{planck}: $\Omega_{m0}=0.315$ and $H_0=67.4\ \mbox{km}\,\mbox{s}^{-1}\mbox{Mpc}^{-1}$. With this in mind, we return to the normalization condition, obtaining $\alpha_1+\alpha_2+\alpha_3=1-\Omega_{\mathrm{m,0}} \rightarrow \alpha_2+\alpha_3=1-\Omega_{\mathrm{m,0}}-\alpha_1$ which in turn results as $\alpha_2+\alpha_3=0.675$. To avoid making arbitrary choices for the parameters values, we take $\Lambda$CDM as our fiducial model and use the physically allowed value of $\beta_{1}$ derived above, from Eq. (\ref{eq:cubic}) we calculate $dX/dz$ and compare it with the $\Lambda$CDM model, we impose the slope of $E(z)$ in $z = 0$ to be the same as in the $\Lambda$CDM model, we obtain
\begin{equation}
    \frac{dX}{dz}\bigg|_{z=0}=-\frac{3\Omega_{\mathrm{m,0}}}{3a +2b+c}, \ \ \ \mbox{and} \ \ \ \frac{dX_\Lambda}{dz}\bigg|_{z=0}=3\Omega_{\mathrm{m,0}},
\end{equation}
with $E^{2}_{\Lambda}(z)=\Omega_{\mathrm{m,0}}(1+z)^{3}+1-\Omega_{\mathrm{m,0}}$, therefore, from the above result we see that the combination of coefficients $3a+2b+c$ must be equal to $-1$, thus
\begin{equation}
    3\alpha_3 + 2\alpha_2=-\alpha_1.
\end{equation}
Using the normalization condition, $\alpha_1+\alpha_2+\alpha_3=1-\Omega_{\mathrm{m,0}}$ as well, we arrive at $\alpha_2=3-3\Omega_{\mathrm{m,0}}-2\beta_1=2.035$ and $\alpha_3=-2+2\Omega_{\mathrm{m,0}} + \beta_1=-1.36$. Now, we can find the values for $\beta_i$ that maintain viable physics and also reproduce the cosmic expansion predicted by the $\Lambda$CDM model at late times. So,
\begin{equation}
    \beta_2=\frac{\alpha_2}{H_0^{2}}\approx4.4796555\times10^{-4},\ \ \ \ 
    \beta_3=\frac{\alpha_3}{H_0^{2}}\approx-6.5902110\times10^{-8}.
\end{equation}
By imposing these conditions, the dynamical HDE sector mimics the behavior of a cosmological constant near $z=0$. Although the $\Lambda$CDM model shows some signs of tension, it is important to remember that it still performs extremely well when tested against individual datasets, such as the cosmic microwave background and large-scale structure formation. Additionally, this indicates that the HDE admits an accessible de Sitter phase during a certain period of cosmic evolution, while still permitting departures from the standard behavior scenario. In order to obtain an analytical solution for the Hubble parameter in this holographic model, we employ the well known Cardano's solutions of the cubic equation (\ref{eq:cubic}). Under the usual Liouville's change of variable, $X(z)=u(z)-b/(3a)$, Eq. (\ref{eq:cubic}) becomes 
\begin{equation}
    u^{3}+pu+q=0,
\end{equation}
with real coefficients $p,q$; this form of the equation is commonly known as the {\it incomplete cubic equation}. In this case, the coefficients are $p = -\Delta_0/(3a^2)$ and $q=\Delta_1/(27a^3)$, where we have defined $\Delta_0 \equiv b^2 - 3ac$ and $\Delta_1 = 2b^3 - 9abc + 27a^2 d$, then according to Cardano's method, the real solution for the cubic equation is
\begin{equation}
    u(z)=\sqrt[3]{-\frac{q(z)}{2}+\sqrt{\Delta(z)}}+\sqrt[3]{-\frac{q(z)}{2}-\sqrt{\Delta(z)}},
\end{equation}
where $\Delta(z) = \left(q(z)/2\right)^2+\left(p/3\right)^3 > 0$. The solution for the Hubble parameter is given as follows
\begin{equation}
 H(z) = H_0 \sqrt{X(z)}, \label{eq:solution}
\end{equation}
and $X(z)=u(z)-b/(3a)$, as previously commented. The physical solution corresponds to the continuous positive root with $X(0)=1$. In Figure (\ref{fig:Hzhde}), we show the behavior of the analytical solution for the Hubble parameter taking into account the values obtained in our analysis for the constants appearing in the cosmological model; as can be seen, for $z\simeq 0$, both models are indistinguishable due to the conditions imposed on the values of the parameters $\beta_{1,2,3}$. As mentioned above, this cosmological model maintains the thermodynamic behavior discussed previously and reproduces the $\Lambda$CDM model cosmic expansion at late times. Additionally, we observe that our model satisfies the consistency condition given by the quotient expressed in (\ref{eq:slow}) when Eq. (\ref{eq:coeff}) is used in terms of the redshift, leading to $c'^{2}/c^{2}$. Notice that the coupling constants appearing in the holographic formula were found to be small under certain consistency conditions, then such terms induce small corrections to the dynamics described by the $H^{2}$ term, as expected from the model formulation. 

\begin{figure}[h!]
\centering
\includegraphics[width=0.7\textwidth]{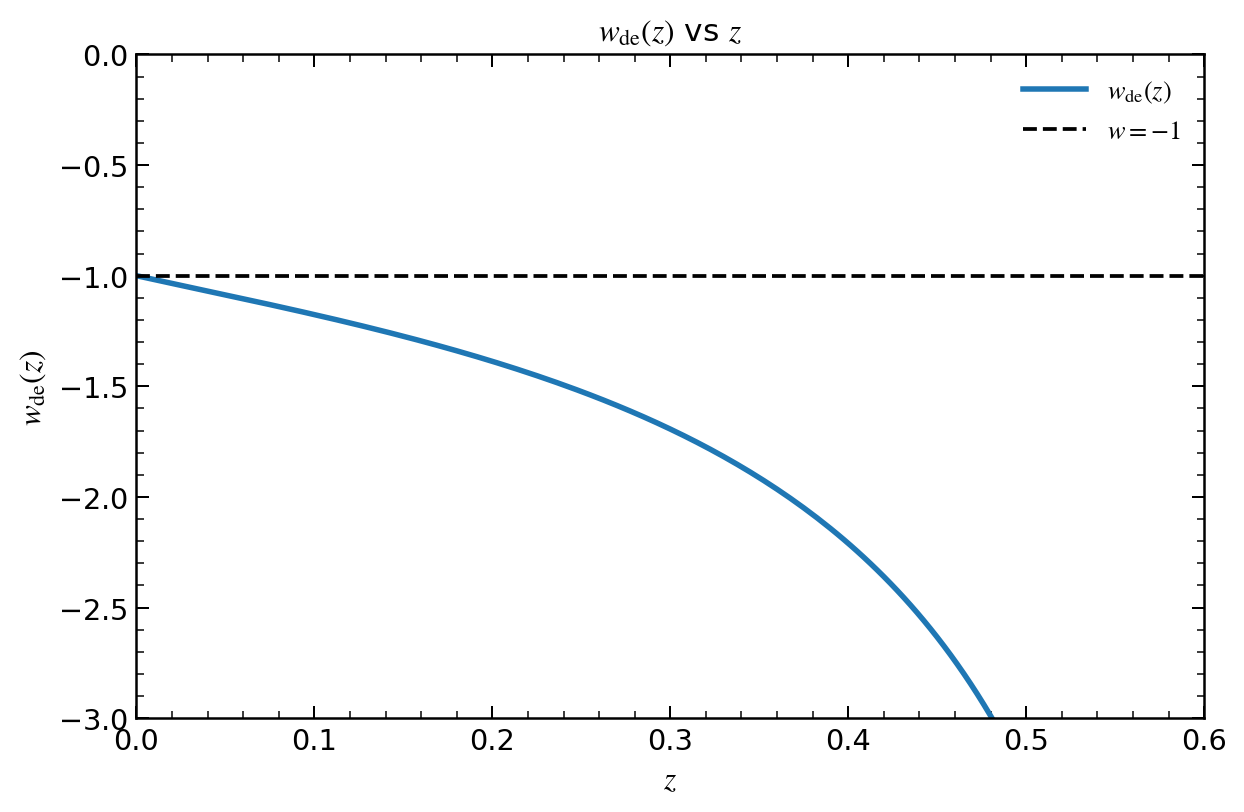}
\caption{Evolution of the dark energy EoS parameter $\omega_{\mathrm{de}}(z)$. The curve remains in the phantom regime for $z>0$ and approaches to $-1$ as $z\rightarrow{0}$, indicating an asymptotic approach to a de Sitter state.}
\label{fig:omega}
\end{figure}

Under the selection of the values found in our analysis for $\beta_{1,2,3}$, together with the solution given in Eq. (\ref{eq:solution}), the energy density (\ref{eq:hdevar}) remains positive.
From the continuity equation for the dark sector,
\begin{equation}
\dot\rho_{\rm de} + 3H\,(1+w_{\rm de})\,\rho_{\rm de}=0,
\end{equation}
we obtain the effective equation of the parameter state
\begin{equation}
w_{\rm de}(z)
= -1 \;-\; \frac{\Omega_{\mathrm{m,0}}(1+z)^3}{\,X(z)\,D\!\left(X(z)\right)}\;
\frac{\alpha_1 X(z) + 2\alpha_2 X^2(z) + 3\alpha_3 X^3(z)}{\alpha_1 X(z) + \alpha_2 X^2(z) + \alpha_3 X^3(z)} ,
\label{eq:wz_general}
\end{equation}
where we have defined $D(X) = 3\alpha_3 X^2 + 2\alpha_2 X + (\alpha_1-1)$. The evaluation at the present time of the latter expression ($z=0$, $X(0)=1$), reduces to $w_{\mathrm{de,0}} = -1$. In Figure (\ref{fig:omega}) we depict Eq. (\ref{eq:wz_general}) in terms of $z$. Consistently with the analysis developed during the search for real analytical solutions for the Hubble parameter, the parameter state approaches $-1$ as $z$ tends toward $0$. However, the model evolves from phantom evolution to a de Sitter stage, the parameter state of the HDE diverges for a positive value of $z$ around $z=0.5$ (not shown in the plot), exhibiting an early (transient) phantom scenario. Then the Universe evolves from the matter domination era to a phantom stage; this kind of behavior was shown to exist in models where powers of $H$ are used to describe the dark energy sector, see Ref. \cite{cruz}, for example. This is significant, as the recent results released by the DESI collaboration present evidence for evolving dark energy, suggesting the possibility of an early time transition from a phantom regime to a present time non phantom regime; see \cite{desi}. Thus, this cosmological scenario is consistent with the thermodynamic framework derived for the proposed holographic model: the late transition from a phantom regime (super-acceleration) to a de Sitter (accelerated) expansion appears, in the thermodynamic description of the model, as a van der Waals–type phase transition.

\begin{figure}[h!]
\centering
\includegraphics[width=0.6\textwidth]{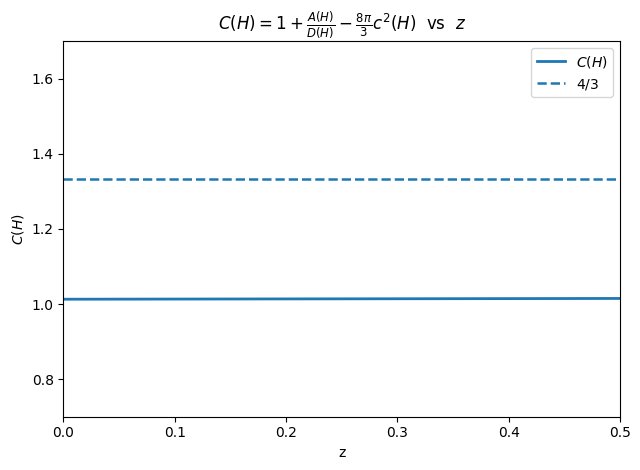}
\caption{Evolution of $C(H)$ in terms of cosmological redshift. We compare against the value $4/3$.}
\label{fig:second}
\end{figure}

To conclude this section, we turn to the generalized second law of thermodynamics (GSL). As stated above, we are modifying the dark sector, including the relevant and marginal deformations in the powers of the factor $H$. This information is enough to determine the evolution of the total entropy of the system
\begin{equation}
    S=S_{h}+S_{M},
\end{equation}
where $S_{h}$ stands for the horizon entropy and $S_{M}$ the entropy contribution coming from the full matter content, baryonic and dark sectors \cite{izquierdo}. Compliance with the GSL implies that $T\dot{S}\geq 0$ once thermal equilibrium has been reached, namely when $T=T_{M}=T_{h}$. In the present case, the temperature is defined as positive. Therefore, it is only necessary to ensure $\dot{S}\geq 0$. As we consider a flat Universe ($k=0$), the apparent horizon entropy is given by 
\begin{equation}\label{B4}
S_h=\frac{A}{4}=\frac{\pi}{ H^2}\quad\Rightarrow\quad \dot S_h=-2\pi\frac{\dot H}{H^3}=3\pi\frac{C(H)}{H},
\end{equation}
where for simplicity in the notation we have defined
\begin{equation}
C(H)=1+\frac{A(H)}{D(H)}-\frac{8\pi }{3}c^{2}(H),\quad A(H)=4\pi \,\left(2\beta_1+4\beta_2H^2+6\beta_3H^4\right),\quad D(H)=3-A(H).
\end{equation}
Now, for the matter content (interior region), using Gibb's relation \cite{izquierdo} one obtains the following result
\begin{equation}
T\,\dot S_{M}=V\dot\rho_{\text{tot}}+(\rho_{\text{tot}}+p_{\text{tot}})\dot V,\qquad
\dot V=-4\pi\frac{\dot H}{H^4},
\end{equation}
with $\rho_{\text{tot}}+p_{\text{tot}}=-\dot H/(4\pi)$ and with $V$ being the volume contained within the apparent horizon, $V=(4\pi \rho_{\text{tot}} R^{3}_{A})/3$, this should be distinguished from the specific volume given by $v=2R_{A}$. So, after some algebraic work, 
\begin{equation}\label{B8}
T\,\dot S_{M}=\left(\frac{\dot H^2}{H^4}+\frac{\dot H}{H^2}\right)\quad\Rightarrow\quad
\dot S_{M}=2\pi\left(\frac{\dot H^2}{H^5}+\frac{\dot H}{H^3}\right)\frac{1}{F(H)},
\end{equation}
with $\dot H=-(3/2)H^2 C(H)$. Thus,
\begin{equation}
\dot S_{M}=\frac{2\pi}{G}\left(\frac{9}{4}\frac{C(H)^2}{H}-\frac{3}{2}\frac{C(H)}{H}\right)\frac{1}{F(H)}.
\end{equation}
In the above expression, the function $F(H)$ corresponds to the dynamical contribution entering in the Hayward-Kodama temperature, that is,
\begin{equation}
T=\frac{\kappa}{2\pi}=\frac{H}{2\pi}\left(1+\frac{\dot H}{2H^2}\right)\equiv \frac{H}{2\pi}\,F(H),\qquad F(H)=1+\frac{\dot H}{2H^2}.
\end{equation}
This function can be expressed as 
\begin{equation}
F(H)=1-\frac{3}{4}C(H).
\end{equation}
Next, putting together (\ref{B4}) and (\ref{B8}) one gets
\begin{equation}
\dot S_{\text{tot}}=\dot S_h+\dot S_{M}=\frac{9\pi}{4}\,\frac{C(H)^2}{H\,F(H)},
\end{equation}
where the following constraints are necessary to ensure the fulfillment of the GSL
\begin{equation}
F(H)=1-\frac{3}{4}C(H)>0\quad\Leftrightarrow\quad C(H)<\frac{4}{3}, \quad H>0.
\end{equation}
These latter conditions are shown in Fig. (\ref{fig:second}), where we have used the values for $\beta_{1,2,3}$ obtained before and our solution (\ref{eq:solution}). Therefore, our HDE model satisfies the GSL at late times.

\section{Final remarks}
\label{sec:5}

In this study, we explored a generalized holographic dark energy model, characterized by a polynomial expansion in the Hubble parameter up to terms of order $H^6$, within a spatially flat FLRW Universe. We examine its thermodynamic behavior with respect to the apparent horizon and investigate its cosmological evolution, comparing it with the conventional $\Lambda$CDM scenario. When linked to the thermodynamics of the apparent horizon, the proposed HDE model displays a rich pattern of critical phenomena. Its resulting equation of state is fully analogous to that of a van der Waals fluid. We established the presence of a first-order phase transition at subcritical temperatures ($T_A < T_c$) and a second-order critical point at $T_A = T_c$. This was verified through an analysis of the $P-v$ isotherms and further supported by the characteristic {\it swallowtail} structure in the Gibbs free energy, indicating the coexistence of two distinct phases.

By restricting the model’s free parameters ($\beta_1, \beta_2, \beta_3$) so that its slope is in agreement with that of $\Lambda$CDM at $z=0$, our framework can replicate the late-time cosmic expansion history of the $\Lambda$CDM scenario, rendering both models indistinguishable around $z \simeq 0$. The corresponding effective dark energy equation of state, denoted $\omega_{de}(z)$, is time-dependent. Although it asymptotically approaches $\omega_{de,0} = -1$ at the present time, it exhibits a late transient phantom regime ($\omega_{de} < -1$) for $z > 0$. This transition from a phantom-dominated phase to a de Sitter-like state on cosmological scales represents the manifestation of the van der Waals-type thermodynamic phase transition inherent to the model.

The holographic model was found to be consistent, satisfying the condition $c'^2(z)/c^2(z) \leq H(z)$ throughout the redshift range considered, and also ensured a positive entropy production. 

In the Appendices (\ref{sec:app1}) and \ref{sec:app2}, it is shown that the model allows for a physically motivated generalization, which could provide a viable framework to address one of the current issues in cosmology, namely the $H_{0}$ tension. Additionally, using the future event horizon ($R_h$) as the IR cutoff also yields an early phantom scenario. A dedicated comparative study between the polynomial $H$-expansion model and the $R_h$ model would be valuable. The terms $H^4$ and $H^6$ were interpreted as higher order corrections from high energy physics. Future theoretical work could aim to derive the small, fitted values of the $\beta_2$ and $\beta_3$ parameters from a more fundamental quantum gravity framework, providing a physical origin for the model's structure. We will report this elsewhere.

 
 \begin{acknowledgments}
M.~Cruz work was partially supported by S.N.I.I. (SECIHTI-M\'exico). J.~Housset acknowledges the warm hospitality of the Physics Faculty colleagues of the Universidad Veracruzana, where part of this work was carried out. The author also acknowledges the financial support of the PAIM-PUCV grant. S.~Lepe acknowledges the FONDECYT grant N°1250969, Chile. J.~Saavedra acknowledges the FONDECYT grant N°1220065, Chile.
\end{acknowledgments}

\appendix
\section{Extended holographic density with Granda-Oliveros term}
\label{sec:app1}
\begin{figure}[h!]
\centering
\includegraphics[width=0.7\textwidth]{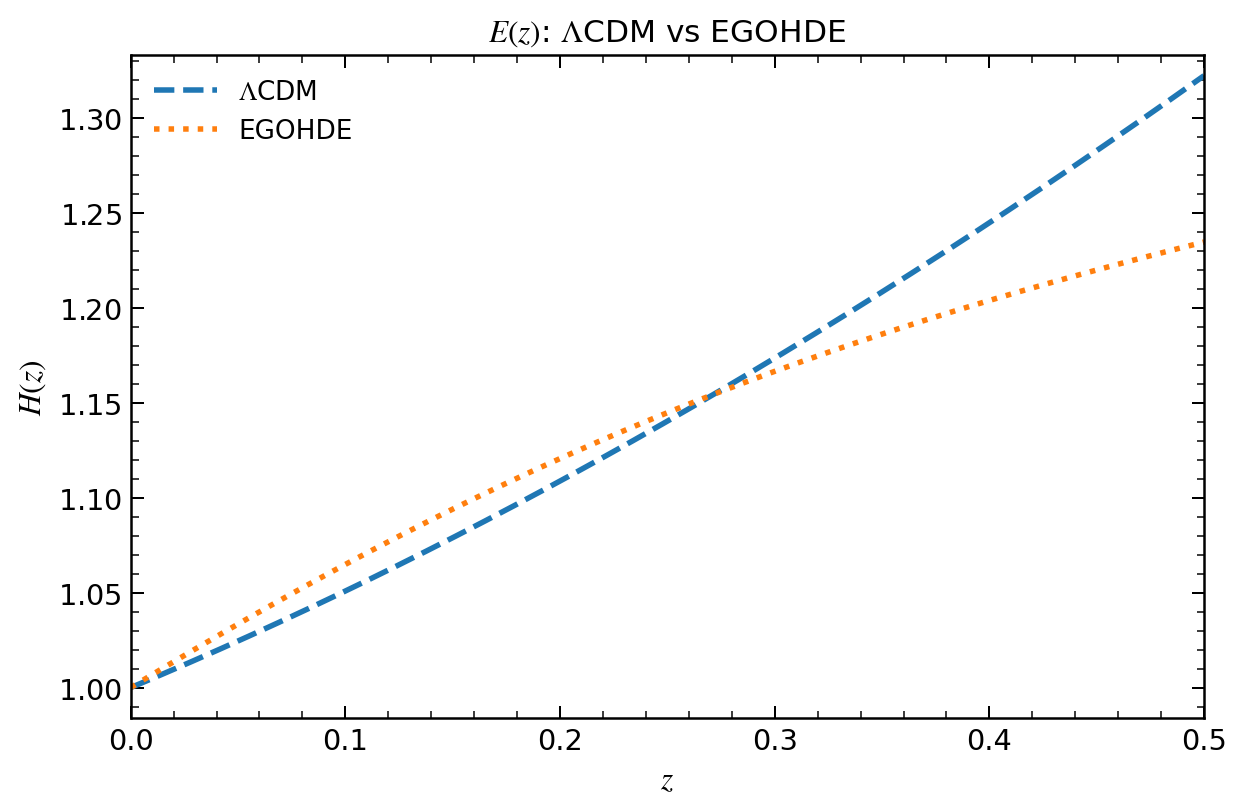}
\caption{$H(z)/H_{0}$ for the EGOHDE model compared to flat $\Lambda$CDM where we take the Planck collaboration results \cite{planck}: $\Omega_{\mathrm{m,0}}=0.315$ and $H_0=67.4\ \mbox{km}\,\mbox{s}^{-1}\mbox{Mpc}^{-1}$.}
\label{fig:ego}
\end{figure}

In this appendix, we show an extension of the holographic model discussed in the context of $P-v$ phase transitions. We study the effects on the cosmological evolution described by our HDE model by an extra contribution to the energy density. We consider the Granda-Oliveros model, which is explicitly given as \cite{Granda:2008dk}
\begin{equation}
\rho_{\rm GO} \;=\; 3\left(\lambda_1 H^2 + \lambda_2 \dot H\right),\label{eq:go}
\end{equation}
where $\lambda_1$ and $\lambda_2$ are free dimensionless parameters. This proposal corresponds to a generalization of the HDE model in which the future event horizon was replaced by the Ricci scalar to avoid causality problems; see Ref. \cite{ricci} for details. It should be stressed that the Granda–Oliveros cutoff yields a geometrically motivated dependence locally on $(H, \dot{H})$, thus departing from conventional HDE formulations. If we repeat the procedure developed before but now incorporate the contribution of (\ref{eq:go}), the modified Friedmann equation reads as follows
\begin{equation}
\alpha_3 E^6(z) + \alpha_2 E^4(z) +
\big(\alpha_1+\sigma_1 -1\big) E^2(z) \;+\; 
\sigma_2 \frac{\dot H}{H_0^2} \;+\;
\Omega_{\mathrm{m,0}}(1+z)^3 \;=\; 0 .
\end{equation}
where we again neglect the contribution of the radiation component since we are interested in late time evolution. Using the relation between cosmic time and redshift,
\begin{equation}
\dot H \;=\; -(1+z)H \frac{dH}{dz}
\;=\; -H_0^2 (1+z)\,E(z)\,\frac{dE}{dz},
\end{equation}
we obtain the following first order differential equation for $E(z)$:
\begin{equation}
\frac{dE}{dz} \;=\;
\frac{ \alpha_3 E^6 + \alpha_2 E^4 +
\big(\alpha_1+\sigma_1 -1\big) E^2 +
\Omega_{\mathrm{m,0}}(1+z)^3 }
{\sigma_2 (1+z) E} ,
\qquad E(0)=1 .
\label{eq:edoE}
\end{equation}
We label this model as {\it extended Granda-Oliveros holographic dark energy} (EGOHDE), equation \eqref{eq:edoE} shows that, unlike the purely algebraic cubic relation of the original holographic model, the inclusion of the Granda-Oliveros term turns the dynamics into a nontrivial first order differential equation in redshift for the Hubble parameter. According to the work of Granda and Oliveros, the interval $\lambda_2 	\,\sim\, 0.5-0.7$ turns out to be compatible with the data and generates a physically reasonable cosmic evolution (transition from decelerated to accelerated expansion). In order to obtain solutions in our model, we will consider $\lambda_2=0.7$ based on the results given in \cite{Granda:2008dk}. Since the value of $\lambda_1$ can be expressed in terms of $\lambda_2$ by means of the normalization condition, as usual, the differential equation for $X$ is solved considering the appropriate values for the model parameters. Equivalently, in terms of $X(z)=E^2(z)$, the Eq. (\ref{eq:edoE}) takes the form:
\begin{equation}
\frac{dX}{dz} \;=\; \frac{2}{\sigma_2}\,
\frac{ \alpha_3 X^3 + \alpha_2 X^2 +
\big(\alpha_1+\sigma_1 -1\big) X +
\Omega_{\mathrm{m,0}}(1+z)^3 }
{1+z} , \qquad X(0)=1 .
\label{eq:edoX}
\end{equation}
Figure \ref{fig:ego} shows how the numerical solution derived from (\ref{eq:edoX}) for the dimensionless Hubble parameter, $E(z)\equiv H(z)/H_{0}$, behaves in comparison with the standard $\Lambda$CDM scenario. The plot clearly demonstrates that the predictions of our model for the Hubble parameter exhibit a slight elevation relative to the prediction of $\Lambda$CDM at low redshifts ($z \lesssim 0.25$). This shift toward larger values suggests that, if $H_{0}$ is treated as a free parameter, a complete statistical analysis (which is not carried out here) might yield a higher estimate for the Hubble constant $H_{0}$ than that predicted by the $\Lambda$CDM model. Consequently, this extended model provides a possible late-time explanation for the existing discrepancy in measurements of the expansion rate, $H_{0}$.

\section{Future event horizon}
\label{sec:app2}
As discussed previously, the existence of critical phenomena in Einstein gravity requires HDE whose dependence on the characteristic length provides a natural way to introduce the use of the geometry of the apparent horizon, especially its radius. In Ref. \cite{cruzlepe}, it was found that the energy density of the form 
\begin{equation}
    \rho_{\mathrm{de}} = 3\beta H^{2n}, \label{eq:hde}
\end{equation}
with $n=1, 2, 3$, supports the existence of $P-v$ phase transitions. We will now describe the aforementioned holographic dark energy in terms of the future event horizon \cite{li}
\begin{equation}
    R_{h} = a\int^{\infty}_{t}\frac{dt}{a}=a\int^{\infty}_{a}\frac{da}{Ha^{2}}. \label{eq:horizon}
\end{equation}
Therefore, the energy density (\ref{eq:hde}) takes the form
\begin{equation}
    \rho_{\mathrm{de}} = 3\beta R_{h}^{-2n}. \label{eq:hde2}
\end{equation}
Taking into account the usual definition for the density parameters $\Omega_{\mathrm{m}}$ and $\Omega_{\mathrm{de}}$, we can write the following from equations (\ref{eq:horizon}) and (\ref{eq:hde2})
\begin{equation}
    \int^{\infty}_{x}\frac{dx}{Ha}=\frac{1}{a}\left(\frac{\beta}{H^{2}\Omega_{de}}\right)^{\frac{1}{2n}},\label{eq:hor}
\end{equation}
where we have defined $x\equiv \ln a$. On the other hand, if the matter sector is described in the usual form $\rho_{\mathrm{m}}=\rho_{\mathrm{m,0}}a^{-3}$, under consideration $a_{0}=1$, then $\Omega_{\mathrm{m}}=\Omega_{\mathrm{m,0}}H^{2}_{0}/(a^{3}H^{2})$ and from the normalization condition $\Omega_{\mathrm{m}}+\Omega_{\mathrm{de}}=1$, we obtain the general expression $\frac{1}{Ha}=\sqrt{a(1-\Omega_{\mathrm{de}})}/(H_{0}\sqrt{\Omega_{\mathrm{m,0}}})$. Inserting the latter result in (\ref{eq:hor}), one is able to write
\begin{equation}
     \int^{\infty}_{x}\frac{dx}{H_{0}\sqrt{\Omega_{\mathrm{m,0}}}}\sqrt{a(1-\Omega_{\mathrm{de}})}=\frac{1}{a}\left(\frac{\beta}{H^{2}\Omega_{de}}\right)^{\frac{1}{2n}},\label{eq:hor2}
\end{equation}
taking the derivative w.r.t. $x$ and after a straightforward calculation, we have
\begin{equation}
    \frac{\Omega'_{\mathrm{de}}}{\Omega^{2}_{\mathrm{de}}}=(3-2n)\left(\frac{1-\Omega_{\mathrm{de}}}{\Omega_{\mathrm{de}}} \right)+\frac{2n}{\Delta}\frac{(1-\Omega_{\mathrm{de}})^{\frac{3n-1}{2n}}}{\Omega^{1-\frac{1}{2n}}_{\mathrm{de}}}\exp\left[-\frac{3}{2}\left(\frac{1-n}{n}\right)x\right],\label{eq:diff}
\end{equation}
where $\Delta$ is a constant defined as $\Delta \equiv \beta^{\frac{1}{2n}}H^{\frac{n-1}{n}}_{0}\Omega_{\mathrm{m,0}}^{\frac{n-1}{2n}}$ and the prime denotes derivatives w.r.t. $x$. Notice that for $n=1$ our expression behaves as the usual holographic dark energy model, $\Omega'_{\mathrm{de}} = \Omega_{\mathrm{de}}(1-\Omega_{\mathrm{de}})\left(1+2\sqrt{\frac{\Omega_{\mathrm{de}}}{\beta}}\right)$ \cite{li,Wang}. Since the general expression (\ref{eq:diff}) depends on the variable $x$, we must solve this differential equation numerically. The derivative w.r.t. cosmic time of (\ref{eq:hde2}) can be computed as $\dot{\rho}_{\mathrm{de}}=-6n\beta R^{-(2n+1)}\dot{R}_{h}$, where $\dot{R}_{h} = HR_{h}-1$ can be calculated from (\ref{eq:horizon}) and according to (\ref{eq:hde2}) we have $R_{h}=(\rho_{\mathrm{de}}/3\beta)^{-1/(2n)}$. Inserting into the continuity equation $\dot{\rho}_{\mathrm{de}}+3H\rho_{\mathrm{de}}(1+\omega_{\mathrm{de}})=0$, we can solve for the parameter state $\omega_{\mathrm{de}}$, which results in
\begin{equation}
    \omega_{\mathrm{de}}=-\frac{3-2n}{3}-\frac{2n}{3}\left(\frac{\Omega_{\mathrm{de}}}{\beta}\right)^{1/(2n)}\left(\frac{H_{0}\sqrt{\Omega_{\mathrm{m,0}}}}{\exp\left[\frac{3x}{2} \right]\sqrt{1-\Omega_{\mathrm{de}}}} \right)^{\frac{1}{n}-1},\label{eq:ev}
\end{equation}
as noted before, for $n=1$ we have $\omega_{\mathrm{de}} = -\frac{1}{3}-\frac{2}{3}\sqrt{\frac{\Omega_{\mathrm{de}}}{\beta}}$, which resembles the usual result for holographic dark energy \cite{li,Wang}. The evolution of the parameter state is known once we solve for $\Omega_{\mathrm{de}}$. Since $x=\ln a$ in terms of the redshift we have $x= - \ln (1+z)$ where we are considering $a_{0}=1$; the following conditions to solve the differential equation are taken into account: $\Omega_{\mathbf{m,0}}=\Omega_{\mathrm{m}}(x=-\ln (1+z)=0) = 0.315$ reported by the Planck collaboration and $\Omega_{\mathrm{de,0}}=1-\Omega_{\mathrm{m,0}}$. The value of $H_{0}$ is again the value of the results of the Planck collaboration.    
\begin{figure}[h!]
\centering
\includegraphics[width=0.45\textwidth]{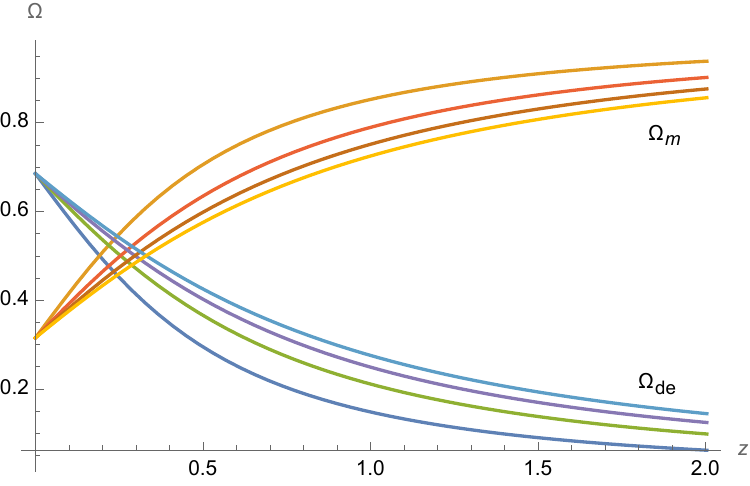}
\includegraphics[width=0.45\textwidth]{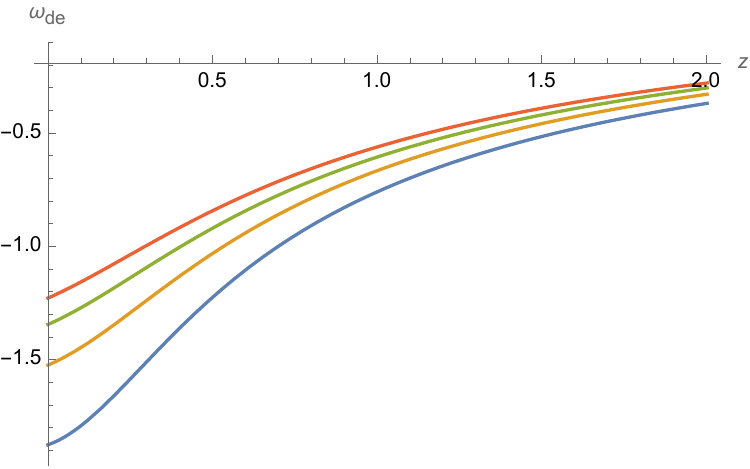}
\caption{Cosmological evolution of the holographic model for $n=2$.}
\label{fig:n2}
\end{figure}
In Figure (\ref{fig:n2}) we show the numerical solution obtained for the differential equation (\ref{eq:diff}) and the parameter state given in (\ref{eq:ev}) for $n=2$ and considering different values for the parameter $\beta$ where $\beta \simeq \mathcal{O}(10^{-4})$. For $n=1$ and $n=3$ similar behaviors as the one shown in the plots can be obtained with $\beta \simeq \mathcal{O}(10^{-1})$ and $\beta \simeq \mathcal{O}(10^{-8})$, respectively; notice that these values for the coupling constants are in agreement with the values obtained in the work for the holographic model. In analogy to the holographic model discussed in the work, this formulation also exhibits the emergence of an early phantom scenario; therefore, both approaches provide similar information for the cosmological evolution. The emergence of the phantom regime using the future horizon to describe dark energy was also discussed in \cite{horizons}.   

\bibliographystyle{unsrt}

\end{document}